\documentclass[11pt]{article}
\usepackage{fullpage}
\usepackage{amsmath}
\usepackage{multicol}
\usepackage[utf8]{inputenc}
\usepackage[english]{babel}
\usepackage{indentfirst}
\usepackage[margin=0.75in]{geometry}
\usepackage{graphicx}
\graphicspath{ {fig/} }
\usepackage{float}
\usepackage{multicol}

\usepackage{color}

\usepackage[pdftitle={Hunting Co-operation in the Middle Predator in Three Species Food Chain Model}, pdfauthor={Nishchal Sapkota},
colorlinks=true, pdfstartview=FitV, linkcolor=blue, citecolor=blue, urlcolor=blue]{hyperref}

\newtheorem{theorem}{Theorem}[section]

\title{\textbf{Hunting Co-operation  in the Middle Predator in Three Species Food Chain Model}}
\author{Nishchal Sapkota $^{1,2}$, Rimsha Bhatta $^3$, Phillip Dabney $^2$\\
Advisor: Dr. Zhifu Xie $^1$ \\
$^1$ School of Mathematics and Natural Sciences \\
$^2$ School of Computing Sciences and Computer Engineering\\
$^3$ School of Polymer Science and Engineering \\
The University of Southern Mississippi\\
Hattiesburg, Mississippi 39406\\
The United States of America \\
email: nishchal.sapkota@usm.edu, rimsha.bhatta@usm.edu, \\phillip.dabney@usm.edu, zhifu.xie@usm.edu
}
\date{}

\begin{document}
\setlength{\parskip}{0cm} 

%
%

\maketitle
%
%

\begin{abstract}

We proposed a three-species food chain model with hunting co-operation among the middle predator. In this model, third species prey on the middle species and the middle prey on the first species. The hunting cooperation among the middle predator affects interestingly on the numbers of both the predators and the prey. We examined the linear stability of the model theoretically and numerically . We conducted the two-parameter numerical analysis to check the long-term behavior and the change in the number of the species with respect to hunting co-operation. Our findings supported the postulates from the two species food chain model with hunting co-operation. \\


\end{abstract}

%
%


%
%

\section{INTRODUCTION}

One of the ways to describe the ecological system is through a food chain/web model between predator and prey. In mathematics, the classical predator-prey model has been analyzed by constructing a mathematical model to help us solve the fundamental biological problems. The interaction between the three species on a linear food chain model is well studied in a three-dimension \cite{LTBT}. One of the natural and realistic cases of a simple food chain model is the one where the top predator is a generalist and the middle predator is a specialist \cite{LTBT}. A generalist predator is defined as the predator who can survive even in the absence of its special food, whereas a specialist predator is the one which can survive only in the presence of special or favorite food. The specialist predator dies in the absence of specific food. A special parameter, $\alpha$, is introduced in the model. $\alpha$ is the hunting cooperation under which, the variation in dynamical behavior or the chaotic behavior is studied. Originally, this had been studied by Upadhyay et. al \cite{UPD} depending on the various functional response. \\
    
Animals hunt together in groups under the division of labor and specified roles \cite{HCP}. This kind of behavior involves two or more animal-eating species, successfully capturing a common prey, which helps all the involved individual to minimize the cost of time and effort than when alone. Here, the three species food chain model is combined with the hunting co-operation to see the changes in the long term behavior of the population. Thereby, the variation in dynamics brought by the initial value condition of hunting cooperation under several constant parameters is studied and their long -term behavior is noted. The mainstream of the paper is to focus on the effect of hunting cooperation in three species models. Numerical methods to solve the system of differential equations, six-stage fifth-order Runge-Kutta method, is implemented in MATLAB. \\
        
    The relationship between hunting cooperation and growth rate is modeled by keeping all the other parameters constant. Our main goal is to analyze how the different intensities of hunting co-operation bring changes in the population density, survival of species and the stability of the ecosystem dynamics. 

\subsection{Hunting Co-operation in Three Species}
The illustrated three species food chain model is the combination of a three species model and Hunting cooperation. It is modified from the classical Upadhyay-lyengar-Rai model by adding hunting cooperation in the specialist predator \cite{HCP}.
	\begin{equation} \label{eq1}
	\begin{cases}
	\frac {d u_1}{dt}=u_1(a_1-b_1u_1-\frac{w_0(1+\alpha u_2)u_2}{(1+\alpha u_2)u_1+D_0}),\\ 
	\frac {d u_2}{dt}=u_2 (-a_2+\frac{w_0(1+\alpha u_2)u_1}{(1+\alpha u_2)u_1+D_1}-\frac{w_2 u_3}{u_2+D_2})\\ 
	\frac {d u_3}{dt}=u_3^{2}(a_3-\frac {w_3}{u_2+D_3})\\
	\end{cases}
	\end{equation}
	
Here, $u_1$, $u_2$ and $u_3$ denote the three species, prey, middle predator and top predator respectively. $\alpha$ denotes the hunting cooperation. $b_1$ is the intra-species competition in the prey. $a_1$ denotes the growth rate of $u_1$, $a_2$ denotes the rate at which $u_2$ dies out in absence of $u_1$ and $u_3$, and $a_3$ is the growth rate of $u_3$ through sexual reproduction. $w$ is the maximum value which per capita reduction rate can attain. In other words, it is the maximum predation that can occur among the species. $D_0$ and $D_1$ signifies the maximum limit to which environment provides protection to the prey $u_1$. $D_2$ is the value of $u_2$ at which its per capital removal rate becomes $\frac{w_2}{2}$ and $D_3$ depicts the residual loss in $u_3$ in absence of its favorite food $u_2$  \cite{UPD}. 

Analysis of this system is an interesting problem as hunting co-operation has barely been analyzed in three species model. This will also help us draw parallels with the similar work done in two species model by Teixeira et.al \cite{HCP}. Furthermore, we could extend the dynamics of three species model explained in Haile and Xie's work \cite{LTBT}.
%
%

\section{STABILITY ANALYSIS}
\subsection{Equilibrium Points} \label{s2}
Rewriting the system of equations (\ref{eq1}) in product form and setting them equal to 0.
	\begin{equation} \label{eq2}
	\begin{cases}
	u_1G_1(u_1,u_2,u_3):=u_1(a_1-b_1u_1-\frac{w_0 u_2}{u_1+D_0})=0\\
	u_2G_2(u_1,u_2,u_3):=u_2(-a_2-b_2 u_2+\frac{w_1 u_1}{u_1+D_1}-\frac{w_2 u_3}{u_2+D_2})=0\\
	u_3^{2}G_3(u_1,u_2,u_3):=u_3^{2}(a_3-\frac {w_3}{u_2+D_3})=0\\
	\end{cases}
	\end{equation}
	
Upon solving (\ref{eq2}) we get the following equilibrium points: \\
$u^{[1]}=(0,0,0)$, \\
$u^{[2]}=(\frac{a_1}{b_1},0,0)$,\\
$u^{[3]}=(u_1^{+},u_2,u_3^{+})$, \\
$u^{[4]}=(u_1^{-},u_2,u_3^{-})$, \\ 
where, \\ 
$u_1^{\pm}=\frac{(a_1 H-b_1 D_0)\pm \sqrt{(b_1 D_0-a_1H)^{2}-4b_1H(w_0 u_2 H-a_1 D_0}}{2b_1H}$, \\
$u_2=\frac{w_3}{a_3}-D_3$, \\
$u_3^{\pm}=\frac{(u_2+D_2)}{w_2}\left[\frac{w_1 u_1^{\pm} H}{(u_1^{\pm}H+D_1)}-a_2 \right] $, \\ \\
for $H= (1+\alpha u_2)$.
	 
However, for some choices of parameters, $u_1^-$ and $u_3^-$ yield either negative or non real values, hence, we avoid $u^{[4]}$ in our analysis. Similarly, $u^{[1]}$ is trivial, and $u^{[2]}$ doesn't offer much either. Hence we will focus more on $u^{[3]}$.

\subsection{Linear Stability for Equilibrium $u^{[3]}$}

To study the linear stability of the model, the Jacobin matrix ($J$) is calculated by taking the partial differentiation of system of equations (\ref{eq2}), with respect to $u_1, u_2, u_3$ respectively.
			\[ 
			J= \begin{bmatrix}
			J_{11}&&J_{12}&&J_{13}\\
			J_{21}&&J_{22}&&J_{23}\\
			J_{31}&&J_{32}&&J_{33}\\
			\end{bmatrix} 
			\]		
			
Where, \\
$J_{11}= G_1 +u_1[-b_1+ \frac {w_0 u_2(1+\alpha u_2)^{2}}{[(1+\alpha+u_2)u_1+D_0]^{2}}]$, \\ 
$J_{12}= -u_1[ \frac {w_0(u_1+D_0)+\alpha w_0[2(u_1+D_0)u_2+\alpha u_1 u_2^{2}]}{[u_1+D_0+\alpha u_1 u_2]^{2}}]$, \\ 
$J_{13}=0$ \\ 
$J_{21}=u_2[\frac{D_1 w_1 (1+\alpha u_2)}{[(1+\alpha u_2)u_1+D_1]^{2}}]$, \\ 
$J_{22}=G_2+u_2[\frac{\alpha u_1 w_1 D_1}{[(1+\alpha u_2)u_1+D_1]^{2}}+\frac{w_2 u_3}{[u_2+D_2]^{2}}]$, \\ 
$J_{23}=-u_2[\frac{w_3}{u_2+D_2}]$, \\ 
$J_{31}=0$, \\ 
$J_{32}=(u_3)^{2}[\frac{w_3}{(u_2+D_3)^{2}}]$, \\ 
$J_{33}=2 u_3 G_3$ \\
			
Also, $J_{33}$ is almost equal to zero for the provided parameters. \\
Thus,
			\[ J= \begin{bmatrix}
			J_{11}&&J_{12}&&0\\
			J_{21}&&J_{22}&&J_{23}\\
			0&&J_{32}&&0\\
			\end{bmatrix} \]

\begin{theorem}
Assuming the inequalities  $J_{11}-J_{22} < 0$ and $J_{11}J_{22}>J_{23}J_{32}$ hold when the positive parameters in model \eqref{eq1} are in a set $\Gamma$, if $u^{[3]}$ is a positive coexistence equilibrium solution and the parameters are in the set $\Gamma$, then it is linearly stable.  
\end{theorem}
	
{\it Proof:} From this matrix, the reduced form of the characteristics polynomial is given by the following expression. \cite{LTBT}
		\begin{equation} \label{eq3}
		p(\lambda)=\lambda^{3}+A_1\lambda^{2}+A_2\lambda+A_3
		\end{equation}
		where, \\
		$A_1=-J_{11}-J_{22}$, \\
		$A_2=J_{11}J_{22}-J_{23}J_{32}-J_{21}J_{12}$, \\
		$A_3=J_{11}J_{23}J_{32}.$ \\
		
Using Routh-Hurwitz criterion for third order polynomials, equation (\ref{eq3}) will have a stable solution if, $A_1 > 0, A_3 > 0$ and $A_1 A_2 > A_3$. The positive terms are $J_{21}, J_{22}, J_{32}$ and the negative terms are $J_{12}, J_{23}$. Therefore, $J_{11}$ has to be negative to satisfy $A_1 > 0$. \\	
	
Provided, $A_3$ is zero for some choices of parameter, our characteristics polynomial is further reduced to a second degree expression of the form: 
			\begin{equation} \label{eq4}		
			p(\lambda)=\lambda^{2}+A_1\lambda+A_2
			\end{equation}	
			
Now, using the Routh-Hurwitz criterion for second order polynomials, equation (\ref{eq4}) will have a stable solution if $A_1 > 0$ and $A_2 > 0$. We have already reached to the conjecture that $A_1>0$. And, for $A_2 > 0$, the condition $J_{11}J_{22}>J_{23}J_{32}+J_{21}J_{12}$ has to be satisfied. \\

Here, the resultant quantity on the right hand side of the inequality is negative as, $J_{23}$ and $J_{12}$ are negative. Because $J_{22}$ is positive, $J_{11}$ is either a positive number or a negative number such that its product with $J_{22}$ is greater than the right hand side.

%
%
\section{Numerical Results and Long Term Behavior}
To observe how hunting co-operation would bring changes to the long term behavior of the system, we use ode45 function in MATLAB to solve the system (\ref{eq1}).Values of the parameters are chosen referring to the standard values as used in reference \cite{LTBT}: $a_2=1, a_3=0.03, b_1=0.05, w_0=1, w_2=0.55, w_3=1, D_0=10, D_1=10, D_2=10, D_3=20$. \\

\begin{figure}[!h]  \label{f1}
\begin{center}
\includegraphics[width=0.7\linewidth]{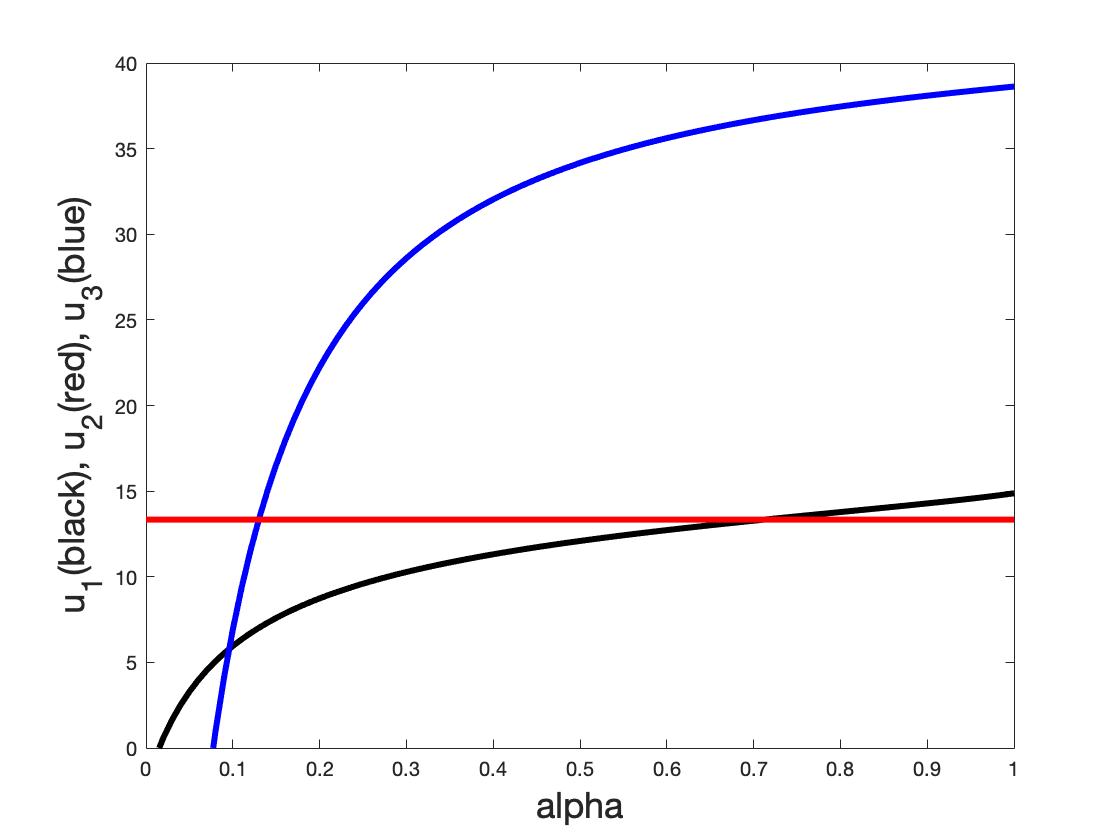}
\caption{Variation of $u_i$ with $\alpha$ ($i = 1,2,3$) at $a_1=1.6$}
\end{center}
\end{figure}
 
 Figure (1) is drawn from the equation of $u_1^+$, $u_2$ and $u_3^+$ in section \ref{s2} with the value of $a_1$ being 1.6. Our concerns for $\alpha$ is within 0 and 1, where any value of $\alpha$ close to 1 indicates the maximum hunting co-operation. The higher value of $\alpha$ could affect the discriminant of $u_1^\pm$, and hence the resultant values for $u_i$ could either be negative or non real values. Here, $u_2$ seems to be unaffected with any degrees of co-operation involved while $u_1$ and $u_3$ seem to be converging to a fixed number as hunting co-operation reaches it maximum value.

\newpage
For the results below, the initial values for $(u_1,u_2,u_3)$ was kept as $(15,13,9)$. We varied the value of $\alpha$ and $a_1$ accordingly, to observe the long term behavior of the system.
\subsection{Cases for $a_1 < 1.5$}

\begin{figure}[!h]		
\begin{center}
\includegraphics[width=0.9\linewidth]{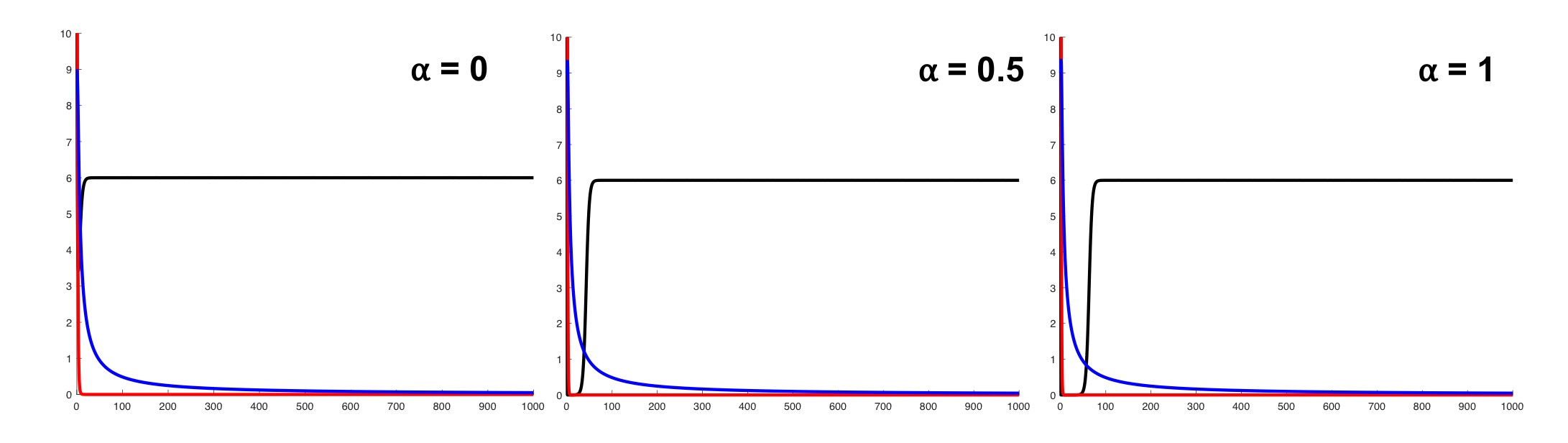}
\caption{($a_1=0.3$) : $u_1$ stable, $u_2$ and $u_3$ extinct}
\includegraphics[width=0.9\linewidth]{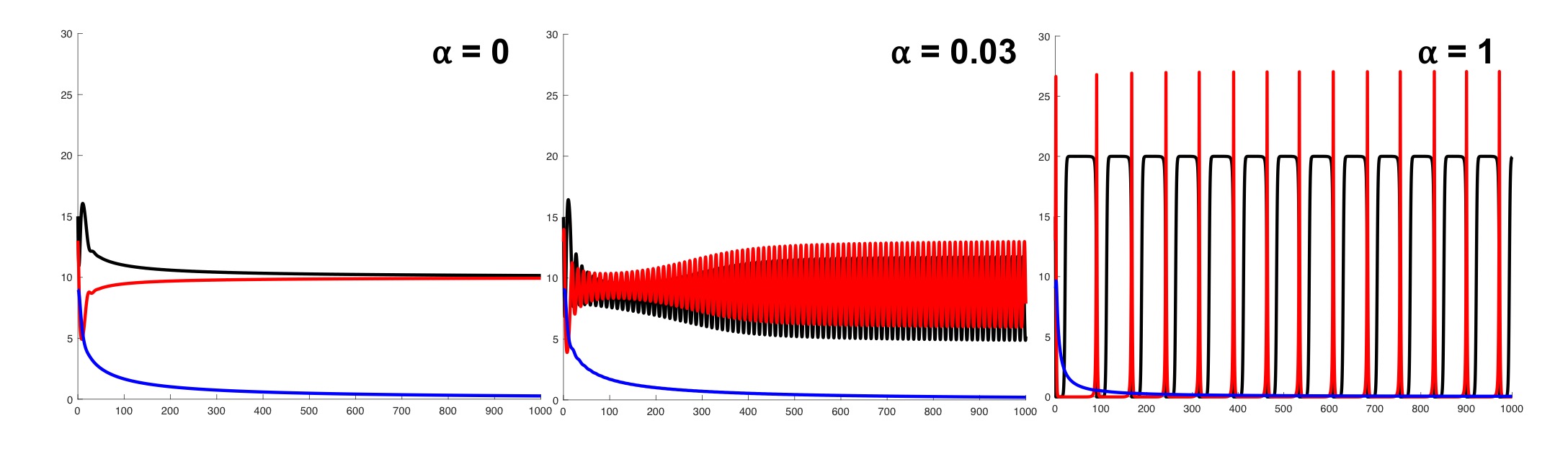}
\caption{($a_1=1$): $u_1$ and $u_2$ stable, $u_3$ extinct}
\includegraphics[width=0.9\linewidth]{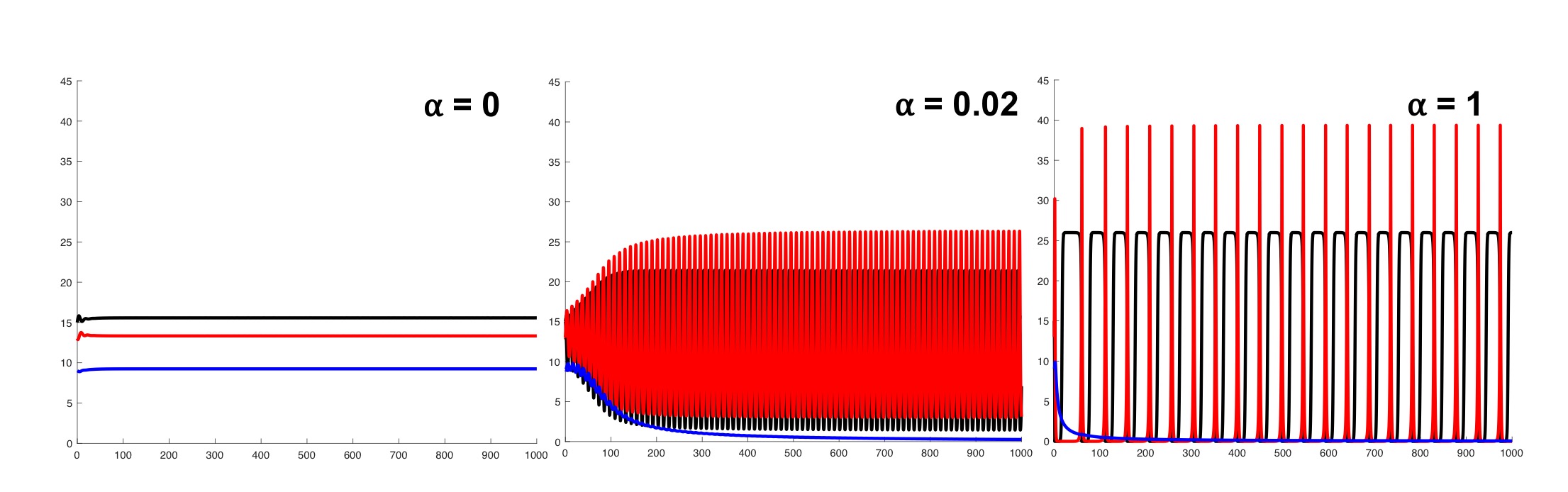}
\caption{($a_1=1.3$): Co-existence}
\end{center}
\end{figure}

In figure 1, changing $\alpha$ did not change the dynamics of the system for small value of $a_1$. However, for higher value of $a_1$, $\alpha$ changed the stability of the system. As the value of $\alpha$ increased, the stable attributes in the graph changed. Stable $u_1$ and $u_2$ was made unstable in figure 2, while the co-existence state of $u_1$, $u_2$, and $u_3$ in figure 3 were also made unstable. In the first two cases, $u_3$ was independent of the change of $\alpha$. However, in the third case, $u_3$ changed from stable co-existence and ultimately seemed to extinct.

\newpage
\subsection{Cases for $a_1 \geq 1.5$}

\begin{figure}[!h]		
\begin{center}
\includegraphics[width=0.65\linewidth]{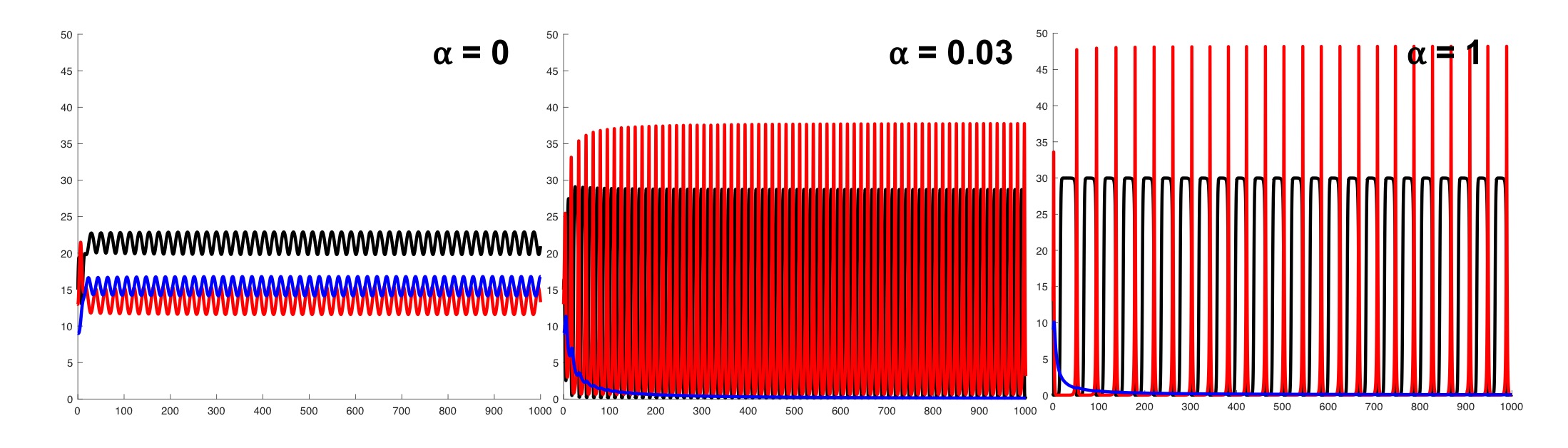}
\caption{($a_1=1.5$): Limit Cycle}
\includegraphics[width=0.65\linewidth]{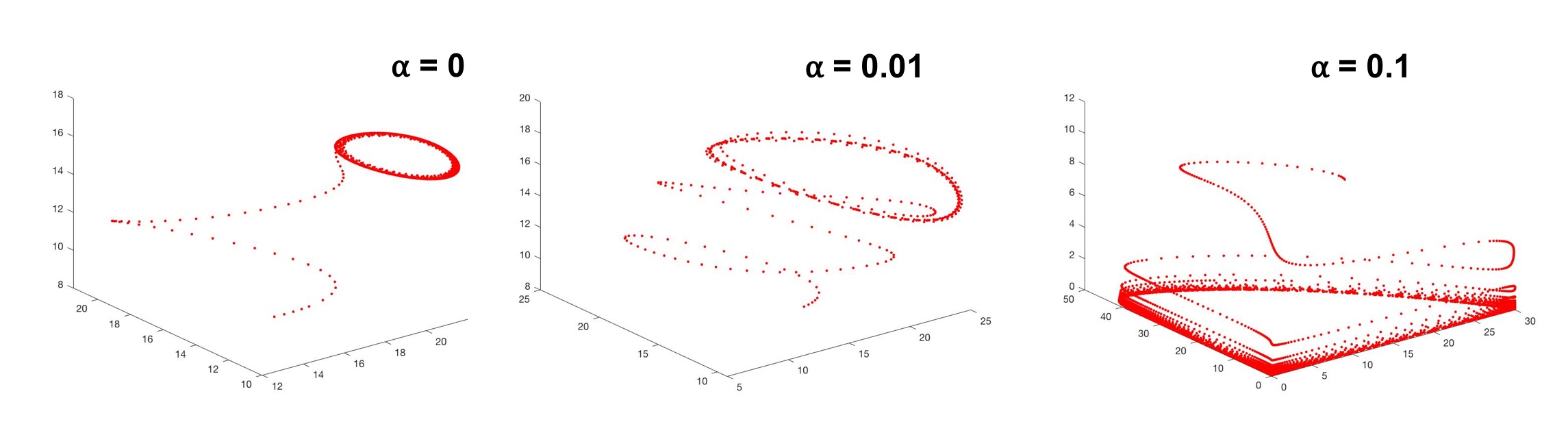}
\caption{($a_1=1.5$): Deformation of Limit Cycle}
\includegraphics[width=0.65\linewidth]{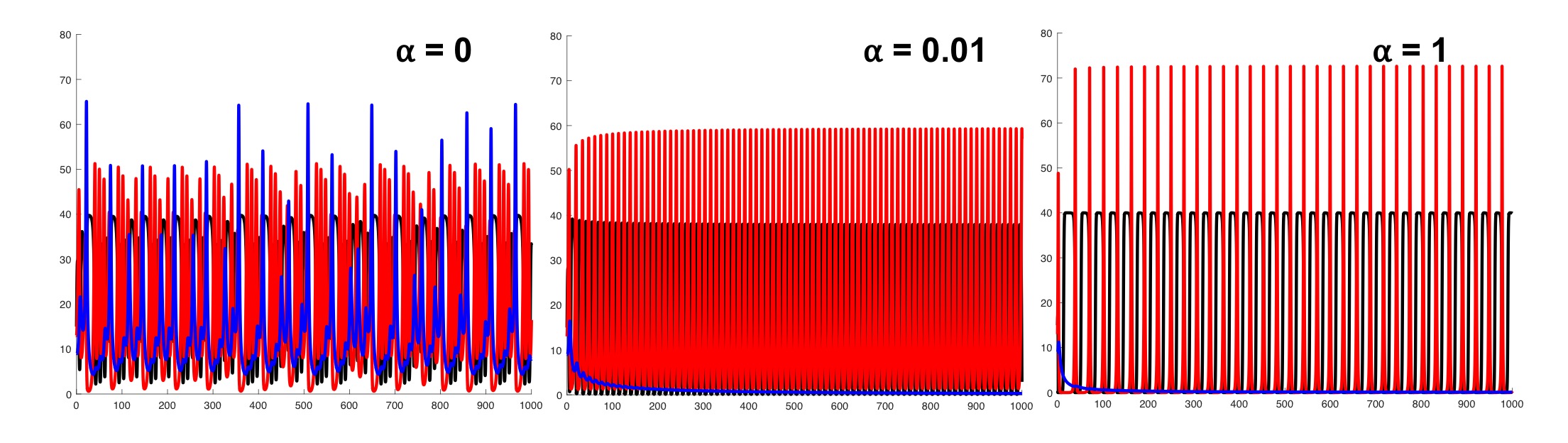}
\caption{($a_1=2$): Chaos}
\includegraphics[width=0.65\linewidth]{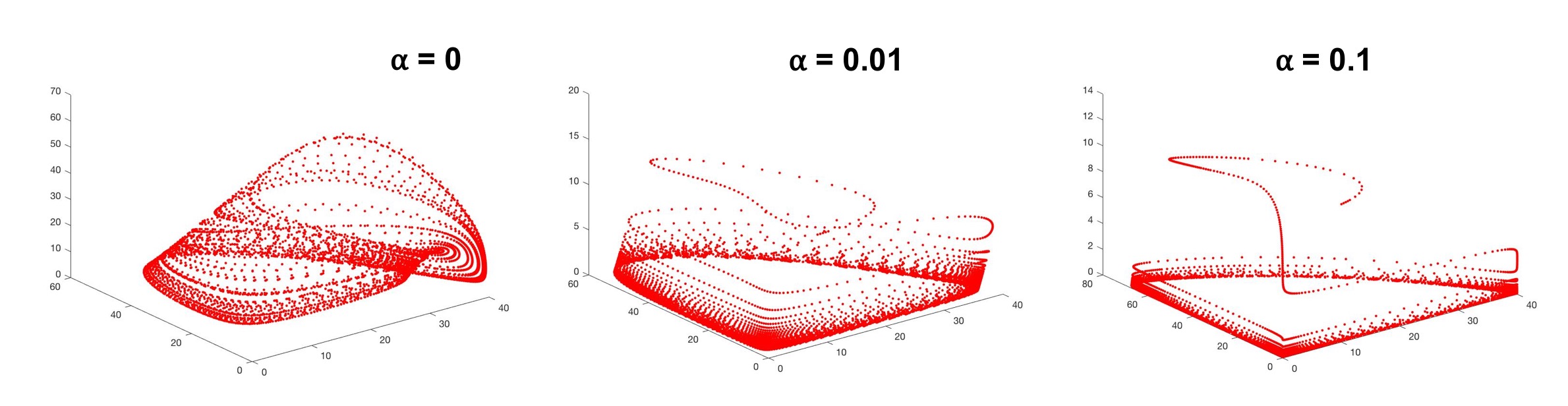}
\caption{($a_1=2$): Deformation of Chaos}
\end{center}
\end{figure}

Limit oscillation is observed for $a_1=1.5$, and chaos is observed for $a_1=2$ in the absence of hunting co-operation ($\alpha=0$). As we increased the value of $\alpha$, both of these were deformed. The amplitude of limit cycle increased as we increased the value of $\alpha$. The deformation was very sensitive even for the small changes in $\alpha$.

%
%
\newpage
\section{CONCLUSION AND FUTURE DIRECTION}
From the numerical results above, we can conclude that hunting co-operation significantly changes the behavior of the prey and the middle predator for higher values of growth rate of the prey. However, the effect of hunting co-operation was minimal when the value of growth rate of prey was smaller. Hunting co-operation leads to the extinction of top predator even for the co-existence equilibrium. It changes the stability of the coexistence equilibrium. Limit cycle oscillations emerges and the amplitude of oscillations increases with hunting co-operation. \\

The results of our work draws parallels to the postulates of hunting co-operation in two species model \cite{HCP}. Further, we aim to study Allee Effects with an anticipation that Alee threshold could occur on the boundary between basin of attraction of co-existence state and basin of attraction of predator-extinction state \cite{HCP}. Identifying the stable and unstable region in graph of top predator and the prey, we could separate the basin of attraction. This could potentially verify that Allee Threshold varies with prey population and for higher prey population, the hunting co-operation leads to smaller Allee Threshold thereby reducing the risk of predator extinction.

%
%
\section{ACKNOWLEDGEMENT}
We are grateful to MAA National Research Experiences for Undergraduates Program in the Mathematical Sciences (NREUP) and Wright W. and Annie Rea Cross Endowed Chair in Mathematics and Undergraduate Research for funding us to participate in this program. We are also thankful to our mentor, Dr. Zhifu Xie  for guiding us through this process. 

%
%


\end{document}